\documentclass[sigconf]{acmart}
\usepackage{color,soul}
\usepackage{subcaption}

\AtBeginDocument{%
  }

\copyrightyear{2026}
\acmYear{2026}
\setcopyright{cc}
\setcctype{by}
\acmConference[ICSE-Companion '26]{2026 IEEE/ACM 48th International Conference on Software Engineering}{April 12--18, 2026}{Rio de Janeiro, Brazil}
\acmBooktitle{2026 IEEE/ACM 48th International Conference on Software Engineering (ICSE-Companion '26), April 12--18, 2026, Rio de Janeiro, Brazil}
\acmPrice{}
\acmDOI{10.1145/3774748.3787601}
\acmISBN{979-8-4007-2296-7/2026/04}

\begin{document}

\title{PriviSense: A Frida-Based Framework for Multi-Sensor Spoofing on Android}

\author{Ibrahim Khalilov}
\email{ikhalil1@jhu.edu}
\affiliation{
  \institution{Johns Hopkins University}
  \city{Baltimore}
  \state{Maryland}
  \country{USA}
}

\author{Chaoran Chen}
\email{cchen25@nd.edu}
\affiliation{%
  \institution{University of Notre Dame}
  \city{Notre Dame}
  \state{Indiana}
  \country{USA}
}

\author{Ziang Xiao}
\email{ziang.xiao@jhu.edu}
\affiliation{
  \institution{Johns Hopkins University}
  \city{Baltimore}
  \state{Maryland}
  \country{USA}
}

\author{Tianshi Li}
\email{tia.li@northeastern.edu}
\affiliation{%
  \institution{Northeastern University}
  \city{Boston}
  \state{Massachusetts}
  \country{USA}
}

\author{Toby Jia-Jun Li}
\email{toby.j.li@nd.edu}
\affiliation{%
  \institution{University of Notre Dame}
  \city{Notre Dame}
  \state{Indiana}
  \country{USA}
}

\author{Yaxing Yao}
\email{yaxing@jhu.edu}
\affiliation{
  \institution{Johns Hopkins University}
  \city{Baltimore}
  \state{Maryland}
  \country{USA}
}


\renewcommand{\shortauthors}{Khalilov, et al.}

\begin{abstract}
Mobile apps increasingly rely on real-time sensor and system data to adapt their behavior to user context. While emulators and instrumented builds offer partial solutions, they often fail to support reproducible testing of context-sensitive app behavior on physical devices. We present \textbf{PriviSense}, a Frida-based, on-device toolkit for runtime spoofing of sensor and system signals on rooted Android devices. PriviSense can script and inject time-varying sensor streams (accelerometer, gyroscope, step counter) and system values (battery level, system time, device metadata) into unmodified apps, enabling reproducible on-device experiments without emulators or app rewrites. Our demo validates real-time spoofing on a rooted Android device across five representative sensor-visualization apps.
By supporting scriptable and reversible manipulation of these values, PriviSense facilitates testing of app logic, uncovering of context-based behaviors, and privacy-focused analysis. 
To ensure ethical use, the code is shared upon request with verified researchers. 

\textbf{Tool Guide:} How to Run PriviSense on Rooted Android \url{https://bit.ly/privisense-guide}

\textbf{Demonstration video:} \url{https://www.youtube.com/watch?v=4Qwnogcc3pw}

\end{abstract}


\maketitle
\renewcommand\footnotetextcopyrightpermission[1]{}

\section{Introduction}

Mobile apps increasingly rely on real-time sensor and system values to drive personalization, functionality, and contextual awareness. Properties such as battery level, ambient temperature, motion, and even system time influence how apps behave \cite{delgado2022survey, Caro-Álvaro2024Gesture-Based, Liu2017Discovering}. For example, apps may reduce their own activity when the battery is low, adapt features based on detected motion, or modify their user interface depending on the context. 

Despite their importance, developers and researchers lack convenient, reproducible ways to manipulate sensor and system signals on real devices \cite{oetzel2014systematic, delgado2022survey, incel2013review, acquisti2017nudges,10.1145/3130941}. Controlled on-device manipulation enables researchers to conduct deterministic tests of edge cases (e.g., low battery, sudden motion, etc.) and systematically probe the causal relationships between sensor inputs and app behavior. 
Emulators often fail to reproduce real-device phenomena such as sensor timing, noise, power management, and hardware-specific behavior.


Prior work has addressed related but distinct challenges: energy profiling and optimization in Android apps \cite{Xu2020An, Bouaffar2021PowDroid, Wang2020An}, on-device debugging in federated learning \cite{Li2022Privacy-Preserving}, secure mobile crowd sensing architectures \cite{Liu2019Energy-Efficient}, and user privacy concerns in augmented reality \cite{Harborth2021Investigating}. 
However, none enables systematic, reproducible spoofing of sensor and system values on physical devices.

This paper fills this gap and presents a new on-device toolkit, PriviSense. It allows users to spoof Android sensor and system values using a lightweight Frida-based setup running entirely on the device. 
Without modifying the apps themselves, it can inject fixed values (e.g., accelerometer, gyroscope output, battery level) into multiple real-world apps for testing and observation.

In the accompanying demo, we show how PriviSense robustly spoofs sensor and system values across five Android apps. Root access enables full control over app-visible sensor and system signals that are otherwise inaccessible on non-rooted devices. In our demonstration, we inject fixed spoofed values for clarity and reproducibility, though PriviSense supports scriptable time-varying signals. This includes time-varying motion sensor spoofing (e.g. accelerometer, gyroscope, step counter, etc.) to mimic physical movement, as well as modifications to system properties (e.g., battery level, system time, and device metadata, etc.) to emulate specific contexts.

\section{Challenges and Targeted Use Cases}

Sensor and system inputs play a critical role in mobile app behavior, yet developers and researchers lack reliable tools to manipulate them on physical devices in a controlled and reproducible manner. Existing solutions, such as emulator environments, logging frameworks, or app-level rewrites, face significant limitations \cite{Caro-Álvaro2024Gesture-Based, delgado2022survey, Wang2025Channel, Luo2020A}. For instance, emulators often fail to replicate hardware-specific behavior, timing variability, power-management interactions, or vendor-specific HAL implementations \cite{androidHAL}. Rewrites of apps require access to source code and introduce maintenance and deployment burdens. Log-based methods, on the other hand, capture behavior passively but do not allow for causal probing or active manipulation of inputs. These challenges hinder the ability to test edge cases (e.g., fall detection, sudden movement, etc.), reveal context-dependent personalization logic, or analyze privacy-related inferences under simulated conditions.

We identify two core use cases where overcoming these challenges is critical:

\textit{\textbf{Software Testing.}}
Developers often need to validate app behavior under rare or hard-to-reproduce conditions. For instance, motion-activated interfaces may rely on precise sequences of gestures \cite{Caro-Álvaro2024Gesture-Based}; emergency apps may only trigger features under extreme sensor values \cite{incel2013review}; battery optimizations may only activate under critical thresholds \cite{delgado2022survey}. Without on-device manipulation of inputs, such scenarios are difficult to reproduce reliably. A testing toolkit that enables reproducible, scriptable spoofing of sensor and system signals can help uncover fragile behaviors, detect failure modes, and improve robustness across diverse user contexts.

\textit{\textbf{Behavioral Auditing.}} In parallel, mobile apps increasingly use sensor and system data to infer user intent, identity, or context. Prior work has shown that even minimal sensor access (e.g., step count or battery status) can enable profiling or activity tracking \cite{delgado2022survey, mahdad2023emoleak}. However, most existing privacy-enhancing tools focus on limiting access to raw data \cite{delgado2022survey}.

We adopt the term \textit{behavioral auditing} to describe an experimental lens that flips the typical data analysis pipeline: rather than collecting real user data and inferring app behavior, researchers inject spoofed inputs to directly probe how apps respond~\cite{10.1145/3613904.3642363,10.1145/3746059.3747798,10.1145/3733816.3760758}. This method is safer, more reproducible, and avoids exposing real user context. Behavioral audits can illuminate opaque personalization logic, uncover sensitive inferences, or validate privacy protections under adversarial conditions \cite{acquisti2017nudges}. For instance, spoofing system time or step counters can reveal whether apps use inferred activity to alter interface behavior, trigger backend logging, or change service availability. These use cases, spanning testing and behavioral auditing, require tooling that supports fine-grained, scriptable control of runtime context on physical mobile devices.

\section{PriviSense: System Design and Workflow} 

\begin{figure}[h]
  \centering
  \includegraphics[width=\linewidth]{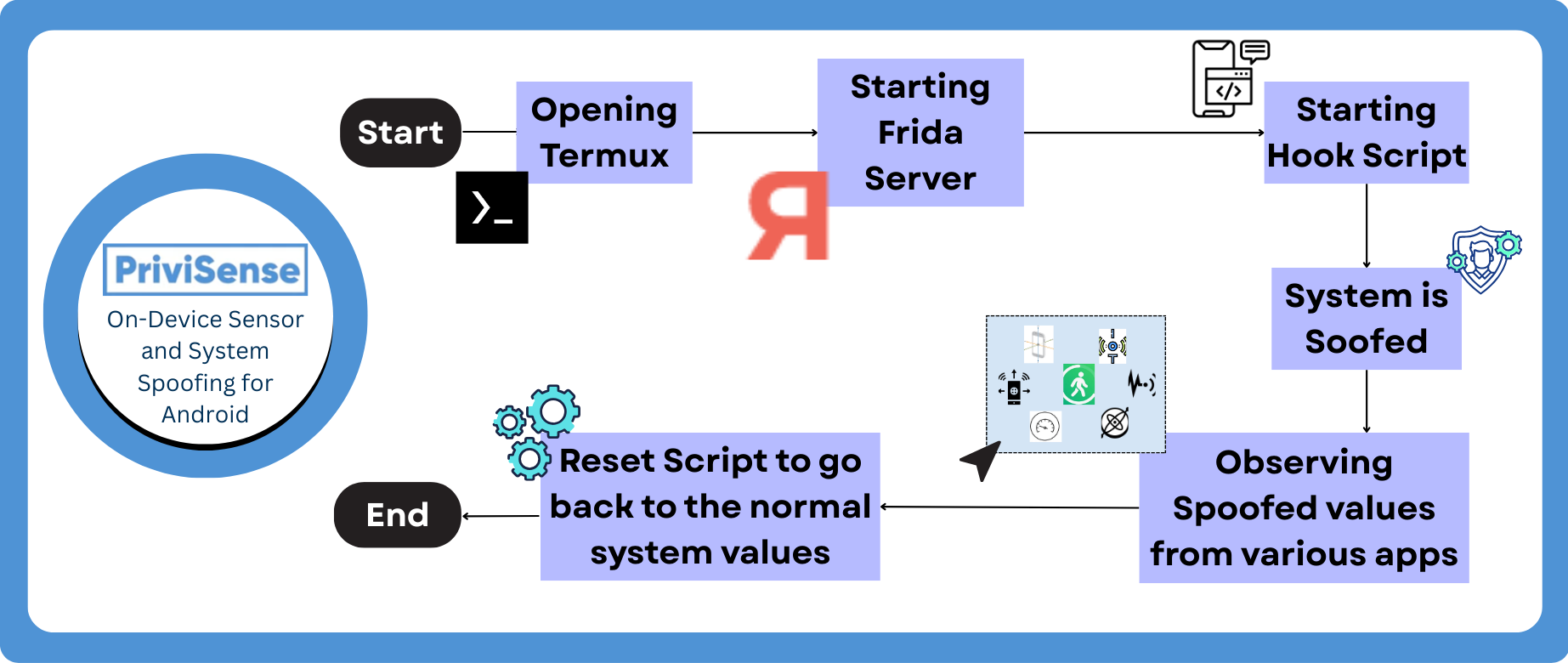}
  \caption{Workflow of PriviSense.}
  \label{fig:workflow}
\end{figure}

At the heart of our approach is a lightweight mechanism for intercepting and modifying runtime API calls that expose system and sensor state to apps. PriviSense achieves this by using Frida \cite{frida}, a dynamic instrumentation framework, to inject JavaScript hooks directly into the Android runtime. These hooks override key system calls related to sensors, battery properties, time settings, and device metadata, allowing apps to perceive a fully spoofed environment, all without requiring an external machine (See: Figure \ref{fig:workflow}).

To support local deployment, we use Termux \cite{termux}, a terminal emulator and Linux-like environment built for Android. However, Android’s native execution model is based on Bionic, a custom C library that does not fully support standard GNU/Linux tooling \cite{bionic}. This limitation means that Python-based Frida tooling and developer dependencies are not directly available on Android devices.

To overcome this, we implemented a custom setup pipeline \cite{privisenseMedium2025} that prepares Termux with the necessary runtime environment. This involves installing Python, compiler toolchains, OpenSSL support, and other dependencies  \cite{openssl}. We then install the correct Frida core devkit for the \texttt{arm64} architecture \cite{arm64} of our device and compile the Python bindings manually. This step is essential because it enables communication between the Termux user shell and the Frida runtime on Android. 

Frida attaches to any running Android process and injects a user-authored JavaScript script. In our case, \texttt{spoof.js} hooks into native Android classes such as \texttt{SensorManager}, \texttt{BatteryManager}, \texttt{SystemClock}, and \texttt{Build}. These hooks override system and sensor calls to return controlled, spoofed values rather than real readings from the device. Although the scripting is written in JavaScript, Frida directly interfaces with Android’s Java layer, allowing seamless runtime modifications without altering the application itself.

\begin{figure}[h]
    \centering
    \begin{subfigure}[b]{0.48\linewidth}
        \includegraphics[width=\linewidth]{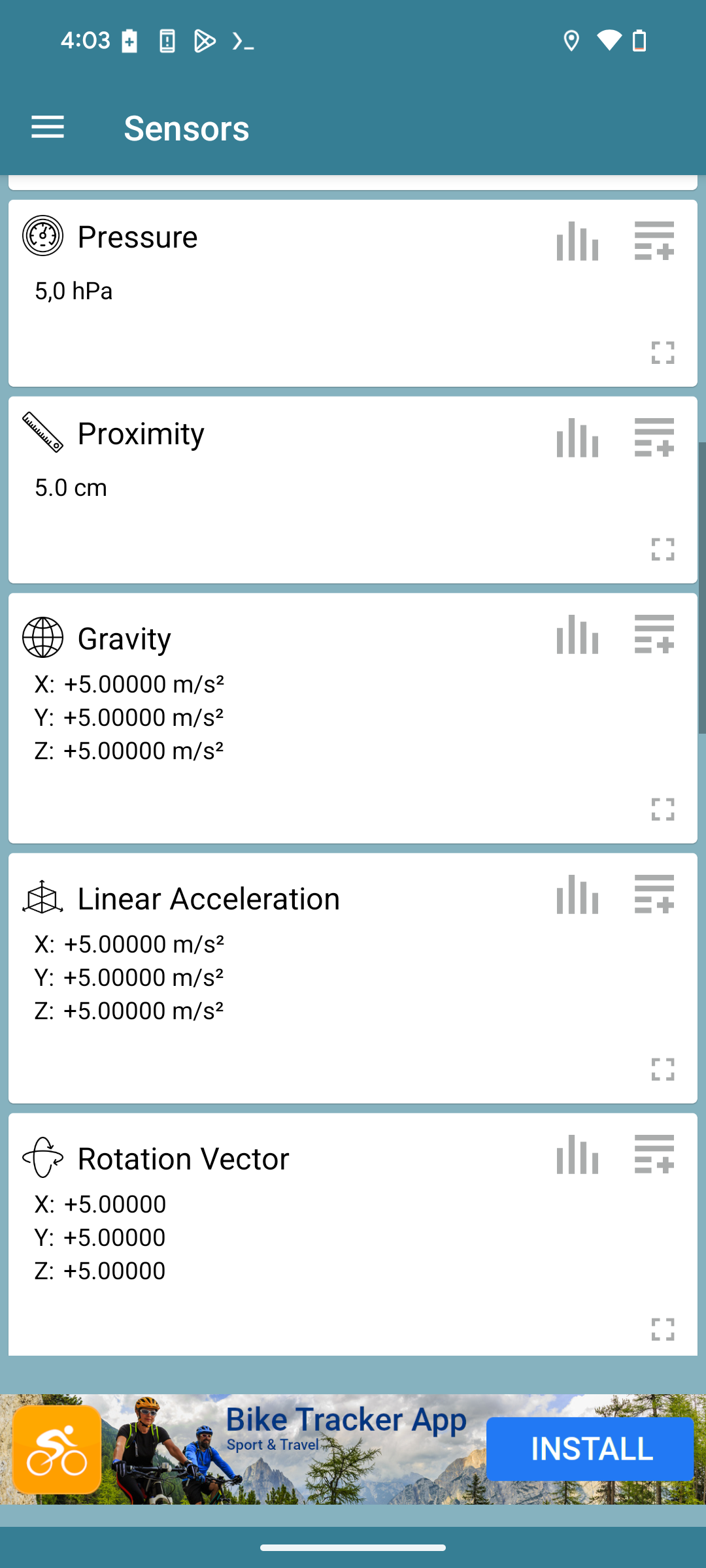}
        \caption{Spoofed values.}
        \label{fig:img1}
    \end{subfigure}
    \hfill
    \begin{subfigure}[b]{0.48\linewidth}
        \includegraphics[width=\linewidth]{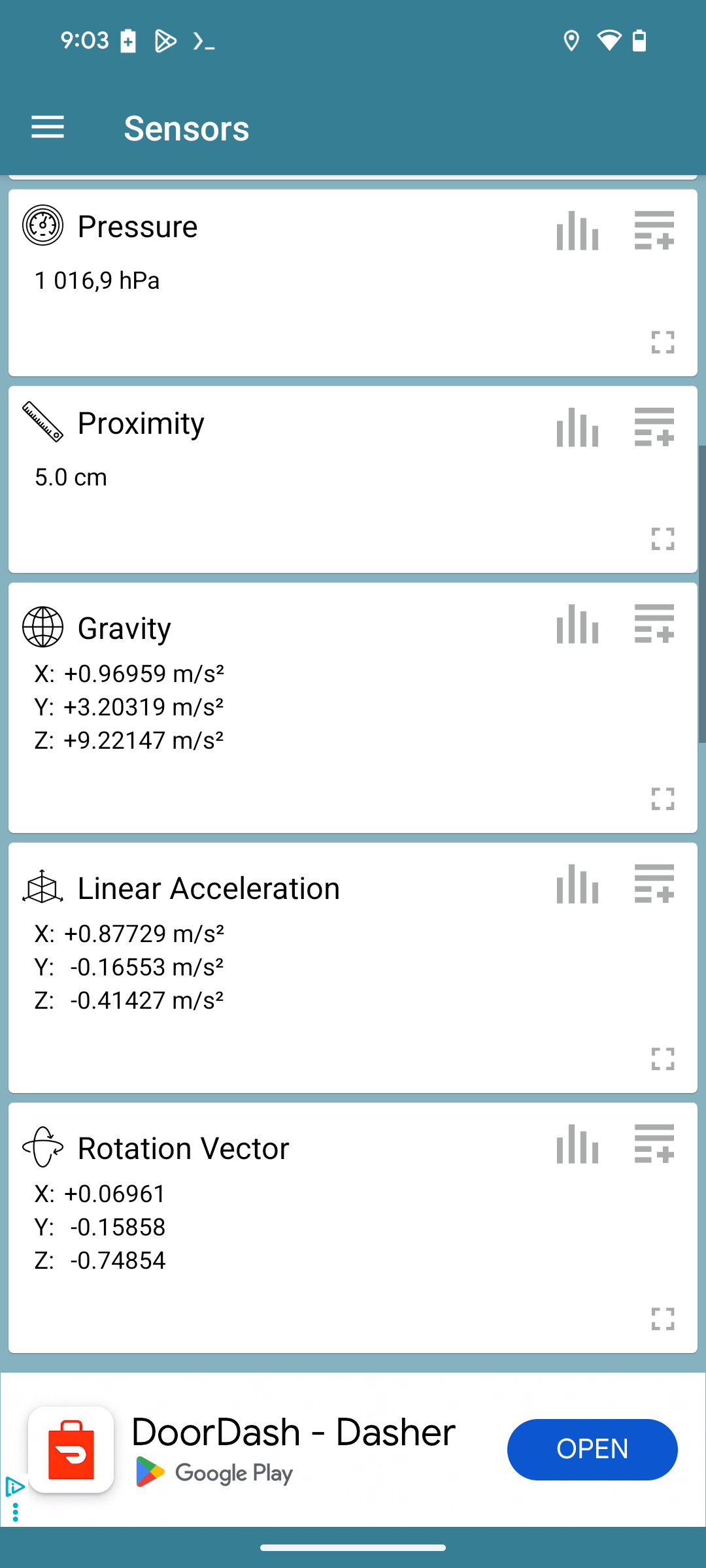}
        \caption{Normal values.}
        \label{fig:img2}
    \end{subfigure}
    \caption{Sensor app output with and without spoofing.}
    \label{fig:side_by_side}
\end{figure}

All execution is initiated directly on the device. The user starts the Frida server with elevated privileges using Termux, launches the script, and opens any target application. The spoofed values are immediately visible across apps that query system state, including diagnostics tools, sensor visualizers, and hardware information dashboards (See: Figure \ref{fig:side_by_side}). 

While our demonstration uses fixed, hard-coded values for clarity and consistency, the underlying infrastructure supports arbitrary spoofing. For instance, it is possible to simulate realistic behaviors such as walking, running, or commuting by injecting time-varying motion data. 
These patterns can be scripted to spoof diverse motion, environmental, and system-level signals.
The system-level spoofing also extends to battery status, temperature, and time. Similarly, users can modify the battery level, system time, or device metadata to simulate different environmental or system states.

This entire pipeline runs on a rooted Android device and does not require emulators, recompiled apps, or firmware modifications. The result is a reproducible and portable testing environment that enables new research and development workflows around personalization, contextual adaptation, and privacy analysis.

\textbf{\textit{Comparison with Related Approaches.}}
Existing tools only partially address the need for controllable, on-device manipulation of sensor and system data. For example, community-developed LSPosed modules such as Motion Emulator can spoof certain values (e.g., motion, location), but are constrained to apps targeting SDK $\leq$30, require system module installations, and lack support for time-varying multi-sensor scripting~\cite{motionemulator}. Other tools focus narrowly on static spoofing of device properties or location, while academic work has primarily targeted energy profiling~\cite{Bouaffar2021PowDroid, Xu2020An}, simulation-based testing, or privacy evaluation through passive logging. These approaches do not offer runtime, app-transparent spoofing across diverse APIs. In contrast, PriviSense uses dynamic instrumentation to inject arbitrary values into sensor and system APIs directly on a physical device, without app rewrites, firmware changes, or external hosts. This enables reproducible behavioral audits and edge-case testing in a unified, scriptable, and portable environment.

\section{Demonstration and Use Cases}
\label{sec:demo}

PriviSense offers a lightweight, reproducible toolkit for spoofing Android sensor and system values directly on a rooted device. Our demonstration showcases how Frida-based JavaScript hooks can inject controlled values into sensor APIs, as well as manipulate battery and system-level identifiers. These capabilities directly support the core use cases outlined in the introduction: (1) software testing under hard-to-reproduce conditions, (2) behavioral auditing of sensor-dependent personalization, and (3) context-triggered behavior analysis in utility and diagnostic apps.

We revisit the use cases through real-world examples. First, in support of software testing, we show how developers can validate mobile applications under rare or context-sensitive conditions. This includes the motivating scenario of asking: \textit{``What would this app do if the phone thinks it is overheating?''} For example, some emergency apps only activate specific features, such as fall detection or crash alerts, when motion sensors exhibit extreme or sudden values. These scenarios are difficult to replicate physically but can be simulated with PriviSense to validate safety-critical functionality. Spoofing motion-related sensors such as accelerometers and gyroscopes also enables testing of activity-specific app behaviors. A fitness app, for instance, may offer different UI flows or trigger background processing depending on whether the user is running or cycling. By simulating those contexts with fixed sensor values, developers can ensure consistent behavior across usage modes.

Second, we revisit behavioral auditing by probing how apps respond to spoofed inputs that resemble real user behavior. This answers questions such as \textit{``How does a fitness app respond to false movement?''} or \textit{``Does step count alone trigger personalization?''} For example, manipulating the step counter reveals how even simple sensor streams can lead to activity tracking without explicit consent. Similarly, changing the perceived device model or Android version might expose how app responses and personalization routines shift based on assumed system identity. These experiments help uncover implicit data collection practices and explore whether apps degrade, adapt, or fail under spoofed contexts.

Third, PriviSense enables exploration of broader app behaviors tied to contextual triggers. We show how spoofing temperature, charging state, or time can affect diagnostic apps or system utilities. In future work, similar techniques could be used to investigate car mode activation, emergency features such as fall detection, or crash response protocols. The key contribution lies not in spoofing alone, but in demonstrating how controlled, repeatable manipulation of sensor and system values can help reveal how apps behave, adapt, or even make decisions based on specific conditions.

Throughout the video demonstration, we launch common diagnostic and sensor apps to visualize spoofed system and sensor signals.
The demo video showcases the consistent manipulation of sensor and system values across different applications, highlighting the tool’s reliability and applicability. At the end of the demo, we restore all the modified sensor and system values. This ensures that the device returns to its original state, allowing clean reruns and preventing unintended carryover between sessions. 

This design makes PriviSense accessible for experimentation and suitable for isolated testing environments, since all spoofing is executed locally on the device without the need for app recompilation or external host machines. The tool is intended only for ethical research and education.

\section{Discussion and Limitations}
\label{sec:discussion}

PriviSense opens up new possibilities for understanding how mobile apps interact with both system and sensor data in controlled experiments. While system properties such as battery level or time can be modified using standard tools, PriviSense expands this capability to runtime sensor APIs (e.g., accelerometer, gyroscope), which are typically harder to manipulate \cite{Lee2021A}. By consolidating these features into a single, scriptable, on-device workflow, the tool enables reproducible testing of app behavior under edge-case scenarios. These include low-battery states, simulated activity patterns, and altered device contexts. 

To prevent potential misuse, we do not release the full codebase publicly. However, we provide fully functional scripts and tooling upon request to verified researchers who demonstrate a legitimate academic or security-related interest. This balances the need for reproducibility with the ethical responsibility to avoid enabling harmful applications of this work.

\textbf{Limitations.} The current implementation requires a rooted Android device, which may not be feasible or desirable in many real-world environments. Some apps may also include runtime defenses that detect the presence of Frida or tampering, causing them to crash or behave inconsistently. These tradeoffs are acceptable for prototyping, debugging, and controlled research studies.

\section{Conclusion}

PriviSense demonstrates how real-time, scriptable sensor and system spoofing can be performed entirely on-device using a lightweight Frida-based framework. 
By spoofing motion sensors, system metadata, and environmental inputs across multiple Android apps, PriviSense enables meaningful analysis of how applications behave in context-sensitive conditions.
Developers can simulate motion-triggered states to test app reactions that are otherwise difficult to induce on demand. 
Researchers can examine how apps personalize features, infer behavior, or adapt based on perceived system identity. 
Unlike emulator-based or app-specific tools, 
PriviSense generalizes to a wide range of use cases and APIs while remaining fully deployable on a single test device.
The accompanying demonstration shows consistently stable manipulation of values across apps, and we share the toolkit with verified researchers to support ethical use and reproducibility. 

\begin{acks}
We thank the anonymous reviewers for their feedback. This work was supported in part by the U.S. National Science Foundation (CNS-2341187, CNS-2426396, CNS-2426397, CNS-2442221, CNS-2426395, CCF-2211428, CMMI-2326378), a Google PSS Faculty Award, and a Meta Research Award.
\end{acks}

\bibliographystyle{ACM-Reference-Format}
\bibliography{bibliography}

\end{document}